\documentclass[prb,showpacs,aps,twocolumn,nofootinbib,superscriptaddress]{revtex4}
\usepackage{graphicx,amssymb,amsmath,natbib}

\begin{document}

\title{Glass transition in the charge density wave system K$_{0.3}$MoO$_{3}$}

\author{D. Stare\v{s}ini\'{c}}

\affiliation{Institute of physics, P.O.B. 304, HR-10001 Zagreb, Croatia}

\author{K. Hosseini}

\affiliation{Experimental Physics II, University of Bayreuth, 95440 Bayreuth, Germany}

\author{W. Br\"{u}tting}

\affiliation{Experimental Physics II, University of Bayreuth, 95440 Bayreuth, Germany}



\author{K. Biljakovi\'{c}}

\affiliation{Institute of physics, P.O.B. 304, HR-10001 Zagreb, Croatia}

\author{E. Riedel}

\affiliation{Laboratory fo Crystallography, University of Bayreuth, 95440 Bayreuth, Germany}

\author{S. van Smaalen}

\affiliation{Laboratory fo Crystallography, University of Bayreuth, 95440 Bayreuth, Germany}

\begin{abstract}

Low frequency dielectric spectroscopy and thermally stimulated discharge measurements of charge density wave
(CDW) system K$_{0.3}$MoO$_{3}$ are presented. Below 80 K two distinct relaxational processes are observed,
which freeze at finite temperatures bearing close resemblance to the phenomenology of the dielectric response
of glasses. We compare our results to the case of o-TaS$_{3}$ in which the glass transition on the level of
CDW superstructure has been recently reported\cite{Star02} and discuss the possibility that it is a universal
feature of CDW systems.

\end{abstract}

\pacs{71.45.Lr, 77.22.-d, 64.70.Pf}

\maketitle

A charge density wave (CDW) is the modulated electronic superstructure that appears in some quasi
one-dimensional systems at low temperatures (for an extensive list of references see \onlinecite{Star02}).
The low energy dynamics of CDW systems is governed by acoustic-like excitations of the phase of the complex
order parameter \cite{phasons}. Apart from the metal-semiconductor transition at a finite temperature T$_{P}$
that is inherent to the CDW formation \cite{CDWbasics}, the most striking features of CDW systems are an
extremely high dielectric constant and nonlinear conductivity at already low electric fields. At the origin
of these phenomena is pinning of the CDW by impurities \cite{impurities} that locks the CDW at the preferred
position, which in turn shifts the onset of CDW sliding to a finite electric field, the so-called threshold
field E$_{T}$. In addition, impurity pinning destroys long range phase coherence and breaks the CDW into
domains.

The CDW dynamics turns out to be strongly temperature dependent, as CDW current \cite{Flem86} and relaxation
frequency \cite{Cava84a,CDWSlowing} both scale with the ohmic conductivity. A hydrodynamic model of the CDW
phase screened by uncondensed free carriers \cite{Sned84,CDWSlowingTheory}, which includes elastic degrees of
freedom, accounted well for these results. However, a qualitative change in the nonlinear conductivity occurs
at lower temperatures\cite{Itkis90,Maed90}, where the CDW displacement is better described by creep
at low or rigid sliding at high electric fields \cite{ZZ93,Ogaw01}, which both neglect elastic degrees of
freedom. It is consistent with the response of CDW in the descreened limit, where the free carrier concentration is too low to 
screen elastic phase deformations \cite{Little88,Nad92} and the intra-CDW Coulomb interaction\cite{Viro93} makes them energetically unfavorable.

Despite numerous evidences that the transition from screened to unscreened response really occurs at finite
temperatures, not enough attention has been given to the transition itself. Our recent paper \cite{Star02} on
wide frequency and temperature range dielectric spectroscopy of the CDW system o-TaS$_{3}$ was devoted to
this particular problem. We have shown that the low frequency relaxational process, that is related to the
dynamics of elastic phase deformations, freezes at finite temperature, and that a secondary process appears
at lower temperatures. As the temperature evolution of both processes bears close resemblance to the
dielectric response of glasses \cite{Glass} it naturally explains the liquid-like to solid-like transition as the glass transition on the level of the CDW superstructure. In this
communication we show that a similar scenario exists in the most widely investigated CDW system
K$_{0.3}$MoO$_{3}$ (blue bronze), which might point to the universality of the glass transition in CDW systems.

DC conductivity $\sigma_{DC}$, I/V characteristics (nonlinear conductivity), AC conductivity $\sigma(\omega)$
and thermally stimulated depolarization (TSD) have been measured in the direction of the highly conducting
axis. Results on two samples did not show any significant difference and we present them for one of
5.7$\times$1.2$\times$0.4 mm$^{3}$ size. We measured $\sigma_{DC}$ and I/V characteristics in four contact
configuration and $\sigma(\omega)$ and TSD in two contact configuration. The contacts were made by clamping
25 $\mu m$ gold wires with silver paste to gold pads evaporated on the crystals.

$\sigma_{DC}$ has been measured between 300 K and 10 K and the I/V curves between 70 K and 10 K. The
$\sigma(\omega)$  has been measured in the frequency range 10$^{-1}\;Hz$ - $10^{7}\;Hz$ at fixed temperatures
between 80 K  and 10 K. We used the frequency-response analyzer Schlumberger (SI 1260) in combination with a broad band dielectric converter (Novocontrol) as a preamplifier. We have verified that the signal amplitude of V$_{ac}$=20 mV kept the response in linear regime. For TSD 
measurements we used Keithley  617 electrometer as both voltage source and current meter. The sample was
cooled in electric fields ranging from 0.4 V/cm to 40 V/cm. After removal of the electric field at low -T the
sample was connected in a short circuit and the discharge current was recorded during the constant rate
heating.

\begin{figure}[b]
\centerline{\includegraphics[width=8.5cm]{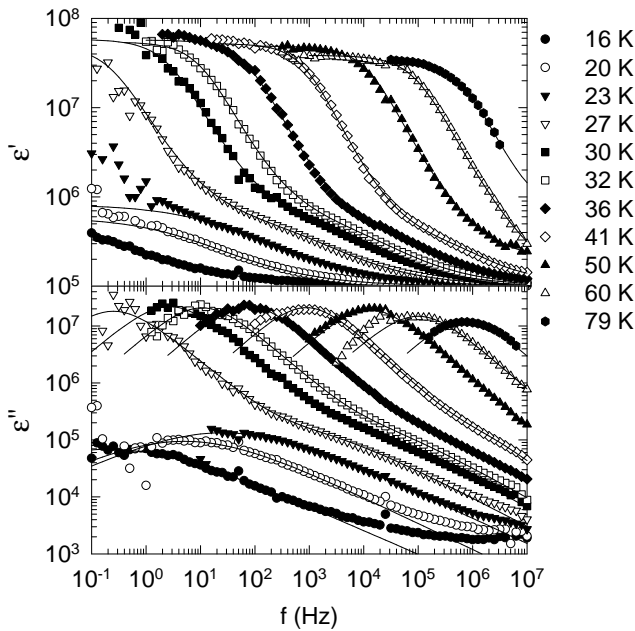}} \caption{\label{fig1} Frequency dependence of the real
($\epsilon$') and imaginary ($\epsilon$'') part of the dielectric response at selected temperatures. Solid
lines represent the fits of the data by CC function.}
\end{figure}

The dielectric function $\epsilon(\omega)$ has been calculated from $\sigma(\omega)$ after subtraction of
$\sigma_{DC}$. The frequency dependence of the real ($\epsilon$') and imaginary ($\epsilon$'') part of the
dielectric function in the units of dielectric permittivity of vacuum $\epsilon_{0}$ is presented in figure
\ref{fig1}. The wide step-like decrease of $\epsilon$' with frequency and the wide maximum in $\epsilon$''
are typical for the overdamped, or relaxational dielectric response. The relaxational processes are
characterized by their amplitude $\Delta\epsilon$, mean relaxation time $\tau$ (or relaxation frequency
$\nu_{0}=1/2\pi\tau$ which gives the position of the maximum in $\epsilon$'') and by the width \textit{w}. In
order to extract these parameters, we have used the modified Debye function with variable width, also known as Cole-Cole (CC)
function:
\begin{equation}
\epsilon (\omega )=\epsilon _{HF} +\frac{\Delta \epsilon }{1+(i\cdot\omega \cdot \tau )^{1/w} }
\end{equation}
where $\epsilon$$_{HF}$ is the high frequency ``base line'' of $\epsilon$'.

From figure \ref{fig1} it is evident that the low frequency dielectric response of K$_{0.3}$MoO$_{3}$ cannot
be attributed to a single relaxation process. The process that dominates at higher temperatures has already
been reported in several papers \cite{Cava84a,Cava84b,Cava85a,Cava86a}, and its features, in particular the
increase of $\tau$ at low temperatures, correspond well to the published data. Due to the extended low frequency window
we have been able to follow the temperature evolution to much lower temperatures and we can see that $\tau$
increases below 30 K without any sign of saturation. This is in agreement with
data obtained from the real time relaxation \cite{Kriz86} and temperature scans at fixed
frequency\cite{Yang91}. However, already at 41 K a high frequency tail develops, which evolves at low temperatures
 into another well defined relaxational process of much lower amplitude. In compliance with the
results of o-TaS$_{3}$ we name the high temperature process primary or $\alpha$ process, and the low temperature one secondary or
$\beta$ process. This is for the first time that the second relaxational process at low temperatures has been
considered and characterized in blue bronze, although signs of the high frequency tail of the primary process
can also be seen in Ref. \onlinecite{Cava85a}. The data have been fitted to two CC functions (Eq. (1)) between 41 K and 27 K, 
corresponding to the coexistence of $\alpha$ and $\beta$ processes, and with a single CC function otherwise. 
The fits are represented with solid lines in figure \ref{fig1}.

\begin{figure}[b]
\centerline{\includegraphics[width=7cm]{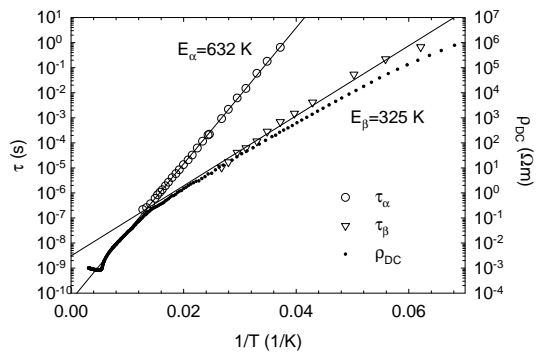}} \caption{\label{fig2} Temperature dependence of the
characteristic relaxation time of $\alpha$ and $\beta$ processes presented in Arrhenius type plot together
with the scaled DC resistivity. Solid lines are fits to the activated behaviour.}
\end{figure}

In figure \ref{fig2} we present the temperature dependence of the mean relaxation times $\tau_{\alpha}$ and $\tau_{\beta}$ 
of both processes, together with the temperature dependence of the DC
resistivity. The slowing down of the $\alpha$ process follows an activated or Arrhenius dependence, with the
activation energy of E$_{a}$=630 K (slightly higher than E$_{a}$=530 K obtained from DC resistivity below
T$_{P}$). Such relation has been observed \cite{Yang91} in K$_{0.3}$MoO$_{3}$, as well as in other
semiconducting CDW systems \cite{Star02,CDWSlowing}. $\tau_{\beta}$ also follows an activated dependence with
about two times lower E$_{a}$=325 K, close to E$_{a}$=320 K obtained from low temperature DC resistivity. In
addition, an extrapolation of $\tau_{\beta}$(T)  to higher temperatures indicates that at about 80 K
the $\alpha$ and $\beta$ processes merge. This kind of temperature evolution is very similar to the one
observed \cite{Star02} in the CDW system o-TaS$_{3}$, but also in glasses in the vicinity of the glass
transition temperature \cite{Glass}. For structural glasses there is a convention that the glass transition
temperature T$_{g}$ is the one at which the relaxation time exceeds 10$^{2}$ seconds, signifying that the
corresponding process becomes so slow that it cannot contribute to the dynamics of the system in the
experimental time window. Based on the extrapolation of the activated fits we obtain T$_{g\alpha}\approx$23 K
for the $\alpha$ process, and T$_{g\beta}\approx$13 K for the $\beta$ process.

\begin{figure}[b]
\centerline{\includegraphics[width=7cm]{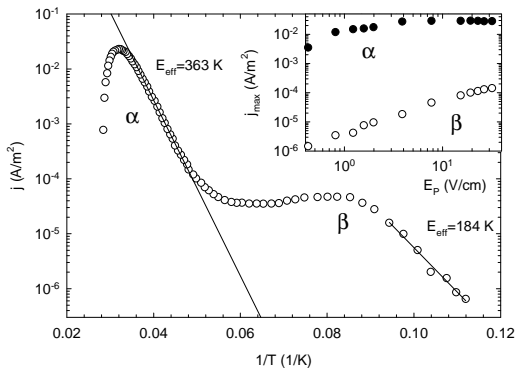}} \caption{\label{fig3} TSD current spectrum recorded at
constant heating rate of 10.5 K/min after cooling from 50 K in electric field of $E_{P}$=4.4 V/cm. Two maxima
correspond to the freezing of $\alpha$ and $\beta$ processes. Solid lines are fits to the activated current
increase. The inset shows the dependence of the current peak values on the polarizing electric field for two
processes.}
\end{figure}

In order to verify that the two processes really freeze at finite temperatures, we have employed TSD
measurements, a general method of investigating low frequency dielectric properties of high resistivity
solids via the study of thermal relaxation effects \cite{TSD}. One TSD spectrum obtained on the same
sample as used for dielectric spectroscopy is presented in figure \ref{fig3}. Two peaks observed
 reveal the freezing of two relaxation processes. The higher one ($\alpha$) is situated at 30 K,
and the lower one ($\beta$) at about 12.5 K, in rough agreement with the dielectric spectroscopy data. The
$\alpha$ peak with very similar features has already been observed before \cite{Cava84} by the TSD method.

From TSD spectra one can obtain several parameters, such as the effective activation energy of the current
increase $E_{eff}$, the position of the maximum $T_{max}$, the maximum current density $j_{max}$ and the
relaxed polarisation $P$. In combination with the known heating rate $h$ and the polarizing electric field
$E_{P}$, it enables the estimation of related dielectric parameters \cite{TSD} such as the effective
relaxation rate $\omega_{eff}=(E_{eff}\cdot h)/(k_{B}\cdot T_{max}^{2})$, the maximum value of
$\epsilon''$  at $T_{max}$ $\epsilon''_{eff}\approx  j_{max}/(\omega_{eff}\cdot E_{P})$ and the
static dielectric constant $\Delta\epsilon_{eff}=P/E_{P}$, as given in  table \ref{tab1}.

\begin{table*}[tbp]
\caption{Effective dielectric data obtained from the TSD spectrum in figure \ref{fig3} and corresponding dielectric spectroscopy data} \label{tab1}
\begin{ruledtabular}
\begin{tabular}{cccccccc}
peak/process & $T_{max}$ (K) & $\omega_{eff}$ (1/s) & $\Delta\epsilon_{eff}$ & $\epsilon''_{eff}$ & T$_{g}$ (K) & $\Delta\epsilon$($\sim$T$_{g}$) & $\epsilon''$($\sim$T$_{g}$)) 
\\ $\alpha$ & 31 & 0.076 & 2.6$\cdot$10$^{8}$ & 7.2$\cdot$10$^{7}$ & 23 & 6$\cdot$10$^{7}$ & 3$\cdot$10$^{7}$ 
\\ $\beta$ & 12.5 & 0.16 & 5.1$\cdot$10$^{5}$ & 9.1$\cdot$10$^{4}$ & 13 & 4$\cdot$10$^{5}$ & 8$\cdot$10$^{4}$ 
\end{tabular}
\end{ruledtabular}
\end{table*}

The parameters estimated for the $\beta$ peak in TSD  very nicely coincide with the parameters of the $\beta$
process obtained from  dielectric spectroscopy, which is not the case for the $\alpha$ peak. In order to
explain this, we present in the inset of figure \ref{fig3} the dependence of the TSD current maxima on the
$E_{P}$ for both peaks. While the $\beta$ peak is in the linear regime in the entire 
range, the $\alpha$ peak approaches saturation already for the lowest applied fields. It has been shown that
both DC bias \cite{Cava84a} and increased signal amplitude \cite{Cava84b} in dielectric spectroscopy lead to
the increase of $\Delta\epsilon$ and $\tau$, therefore it is reasonable that the $\alpha$ process freezes at
higher temperatures and has higher polarizability in the saturated regime.

\begin{figure}[b]
\centerline{\includegraphics[width=7cm]{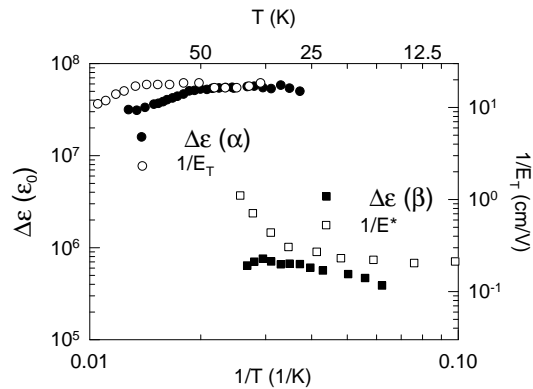}} \caption{\label{fig4} Temperature dependence of
amplitude $\Delta\epsilon$ of $\alpha$ and $\beta$ processes plotted together with the inverse values of two
threshold fields observed in high (E$_{T}$) and low (E$^{*}_{T}$) temperature range.}
\end{figure}

Our results unambiguously show  the existence of two relaxation processes in the low frequency dielectric
response of the CDW system K$_{0.3}$MoO$_{3}$, which is a completely novel feature. Moreover, we have shown
that the temperature evolution follows the same scenario as for the glass transition in o-TaS$_{3}$. The
dynamics of the $\alpha$ process has been thoroughly considered theoretically, and it has been successfully
modeled by the dynamics of the local elastic deformations of the CDW phase \cite{CDWSlowingTheory}. Therefore
the freezing of these excitations reduces the phase space and represents essentially the change in the ground
state in the same way as in structural glasses. This claim can be supported by a number of other experimental
results  that point to the changed properties of K$_{0.3}$MoO$_{3}$ in the range below 40 K  down to 20 K
such as closing of the thermal hysteresis in the DC conductivity, the appearance of a second E$^{*}_{T}$, an
anomaly in proton chanelling, the disappearance of metastabilities in ESR spectra \cite{Dumas93}. Also, the
changes in the lattice parameters \cite{Tian00} of K$_{0.3}$MoO$_{3}$ observed below 50 K indicate that the
glass transition on the level of the CDW superstructure might affect the host lattice as found\cite{Star02a}
for TaS$_{3}$.

In Ref. \onlinecite{Star02} we found that the freezing of the $\alpha$ process in o-TaS$_{3}$ occurs when
there are not enough free carriers to screen efficiently the local phase deformations. We have set a
criterion for the critical density to be about one free carrier per domain of phase coherence. From
extrapolation of the high temperature activated decrease of conductivity we estimate the critical density at $T_g$ for
K$_{0.3}$MoO$_{3}$ to be about 3$\cdot$10$^{13}$ e/cm$^{3}$. The corresponding volume is 3.3$\cdot$10$^{-14}$
cm$^{3}$, which is only by factor of 2 smaller than the phase coherence volume estimated from X-ray
diffraction \cite{DeLa91}. Therefore the same criterion is applicable for K$_{0.3}$MoO$_{3}$ as well.

Although $\Delta\epsilon$ for the $\beta$ process is smaller than for the $\alpha$ process, its value is
still too high to represent the single particle response. We present in figure \ref{fig4} the T-dependence of
$\Delta\epsilon$ for both processes and compare it with the inverse value of two threshold fields measured in
the same sample. Both processes obey approximately the relation $\Delta\epsilon\cdot E_{T}=const$. This has
already been established for the $\alpha$ process in several CDW systems \cite{Wu84}, in corresponding
T-ranges, as well as for the $\beta$ process\cite{Star02} in o-TaS$_{3}$. This close relation to the second
threshold field, as well as the activated increase of $\tau_{\beta}$ folowing $\sigma_{DC}$ support its CDW
origin. Therefore, the $\beta$ process should represent the dynamics of the remaining degrees of freedom of
the CDW after the elastic ones are frozen, i.e. topological or plastic deformations, such as solitons, domain
walls or dislocation loops. These can contribute to the low frequency dielectric response
\cite{Volk93,Lark95} as well as to linear and non-linear conductivity \cite{Lark78,Braz99}. Localized midgap
states have indeed been observed in femtosecond spectroscopy \cite{Dems99}. The approach based on coexistence
of local ("strong") and collective ("weak") pinning\cite{Lark95,Braz99,Tuck88}, or plastic and elastic
deformations of the phase could also naturally account for the coexistence of $\alpha$ and $\beta$ process,
particularly as it has been established that the pinning is locally always strong\cite{Rouz00}.

The temperature evolution of $\alpha$ and $\beta$ processes observed in the two CDW systems o-TaS$_{3}$ and
K$_{0.3}$MoO$_{3}$ share some common properties. In both systems the $\beta$ process splits from the $\alpha$
process at frequencies around 1 MHz and the activation energy E$_{a\beta}$ is about two times smaller than
E$_{a\alpha}$. Also, $\Delta\epsilon_{\alpha}$ is comparable in both systems. However, unlike in
K$_{0.3}$MoO$_{3}$, $\tau_{\alpha}$ in o-TaS$_{3}$ deviates from an activated behaviour \cite{Star02} on
approaching T$_{g}$. Such increase of the effective activation energy close to T$_{g}$ as in o-TaS$_{3}$ is
characteristic of fragile glasses \cite{Cavagna01}, while strong glasses obey an Arrhenius behaviour as in
blue bronze. The differences between strong and fragile behaviour result from the different topography of the
phase space \cite{Grig02}. In CDW systems the phase space is essentially created through pinning, so the
modification of the potential energy landscape reflects the changes in pinning properties. This might finally
bring a completely new light to the problem of pinning in the field of CDWs.

The difference between o-TaS$_{3}$ and K$_{0.3}$MoO$_{3}$ exists also in the amplitudes of the relaxation
processes. While $\Delta\epsilon_{\alpha}$ in blue bronze is almost preserved in the entire T-range, in
o-TaS$_{3}$ it decreases strongly on approaching T$_{g}$. On the other hand $\Delta\epsilon_{\beta}$ is
comparable to $\Delta\epsilon_{\alpha}$ in o-TaS$_{3}$, while in K$_{0.3}$MoO$_{3}$ it is almost two orders
of magnitude smaller. The smaller $\Delta\epsilon_{\beta}$ in K$_{0.3}$MoO$_{3}$ would be consistent with the
smaller number of topological defects and therefore larger phase coherence volume. The same difference in amplitude is also
observed in the low temperature power-law contribution to the heat capacity \cite{Odin01} and in the low
temperature up-turn in magnetic susceptibility \cite{Bilj03}, where both phenomena are again attributed to
topological defects of the phase. Again, fragile glasses typically exhibit a pronounced $\beta$ process
just as in o-TaS$_{3}$, whereas in strong glasses like K$_{0.3}$MoO$_{3}$, there persists only the
"background loss" contribution to the low frequency dielectric response below T$_{g}$ \cite{Wied99}.

In conclusion, we have shown that the glass transition scenario previously seen in o-TaS$_{3}$ exists also in
K$_{0.3}$MoO$_{3}$. This could represent a universal feature of CDW systems as it explains the transition to
the low temperature CDW state and accounts quantitatively for its properties. New approaches to the glass transition
based on the  phase space landscape \cite{Grig02} might help in understanding the pinning properties of CDW
systems and shed light on the differences between o-TaS$_{3}$ and K$_{0.3}$MoO$_{3}$.

\end{document}